\documentclass[twocolumn,prl,showpacs,preprintnumbers,amsmath,amssymb]{revtex4}
\usepackage{graphicx}
\begin{document}
\title{ Collective Flow vs. Hard Processes \\ 
in Relativistic Heavy Ion Collisions}
\author{J. Burward-Hoy and B.V. Jacak} 
\affiliation{Department of Physics and Astronomy, SUNY Stony Brook,
	Stony Brook, NY 11794}
\begin{abstract} 
In relativistic heavy ion collisions, the $p_T$ 
distributions of hadrons reflect the 
transverse motion generated during the collision 
and therefore the collision dynamics.  The moderate 
$p_T$ values measured to date at $\sqrt{s} \approx $ 
20 GeV/nucleon contain a mix of hard and soft 
processes.  The available data from the SPS on $\pi$, 
K, p, and $\Lambda$ are used to constrain the magnitude 
of collective radial expansion.  In $A+A$ collisions, 
the hydrodynamic distributions determined from the measured 
collective expansion are determined up to high $p_T$ and 
are consistent with the $\pi^{0}$ data.  To cleanly observe 
hard processes, measurements at $p_T \ge 4$ GeV/c will be 
required.  Pions produced in $p+W$ collisions clearly 
deviate from hydrodynamic distributions at $p_T > 2$ GeV/c, 
underscoring the differences in the physics between $p+A$ and 
$A+A$ collisions.
\end{abstract}
\pacs{PACS numbers: 25.75.+r}
\maketitle
The possibility of creating a new form of matter, a quark-gluon 
plasma, in ultra-relativistic heavy-ion collisions has generated 
intense experimental and theoretical effort. Transverse momentum 
distributions of hadrons reflect the transverse motion generated 
during the collision and therefore the collision dynamics.  
If there are sufficient secondary scatterings, local thermal 
equilibrium and collective hydrodynamical motion may occur.  
Measured hadronic transverse momentum ($p_T$) distributions 
have been used to search for such collective behavior and 
extract the hydrodynamic parameters \cite{NA44flow,NA49}.

It is, however, the early stage of the collision, when the deconfined 
phase should exist, that is of paramount interest. Hard processes 
create probes for this phase, during the initial collisions between 
the participating nucleons.  For example, particle jets arise from 
the fragmentation of high $p_T$ quarks and gluons produced in inital hard 
scatterings. Study of high $p_T$ hadrons can therefore yield information about 
energy loss of these quarks and gluons as they traverse the hot, 
dense nuclear medium.  Theoretically, hard processes may be 
investigated using perturbative QCD-inspired parton models, tuned to 
reproduce experimental data on nucleon-nucleon (p+p) and 
nucleon-nucleus (p+A) collision systems \cite{hijing}.  Experimentally, 
the challenge is to identify a $p_T$ range accessible with good statistical 
precision, but dominated by jets.

Transverse collective expansion of the colliding system has been 
observed in high energy nucleus-nucleus (A+A) 
collisions \cite{NA44flow,NA49}.  This flow boosts the $p_T$ of 
the hadrons, flattening the spectrum and effectively increasing 
the $p_T$ range where soft processes are important.  The $\pi^0$ 
production in Pb + Pb collisions at 158 GeV/nucleon has been 
successfully reproduced both by partonic models \cite{wangnojets} 
and by hydrodynamics (i.e. soft physics dominated) \cite{WA98flow}.  
This suggests that the complex interplay between hard and soft 
processes may not be unambiguously resolved by a single hadron 
spectrum at $p_T \le$ 3-4 GeV/c.

In this paper we present a systematic study of low $p_T$ hadron 
spectra, using all available data from $\sqrt{s} = $ 17-20 GeV/nucleon 
on $\pi$, K, p and $\Lambda$ to constrain the magnitude of 
collective radial expansion.  The expansion parameters so obtained 
are used to extrapolate the pion spectrum to high $p_T$ for each 
collision system.  The distribution is then compared to the $\pi^0$ 
data in order to determine at what $p_T$ hydrodynamical 
expansion no longer accounts for the particle yield. As initial 
conditions and the surface profile of the expanding collision 
volume are not well constrained by the data, 
we choose the simplest profiles allowed 
by the data, and constrain the parameters using multiple hadronic 
species at transverse mass 
$m_T$ $( = \sqrt{p_T^2 + m_0^2}) - m_0 \le $ 1 GeV.  
We include p+A collisions as well, because the secondary hadronic 
scattering and resulting transverse collective flow should be small.

Collective transverse flow boosts the velocity of the hadrons 
and introduces a hadron mass dependence to the slope of the 
$m_T$ spectra.  At CERN, NA44 \cite{NA44flow} and NA49 \cite{NA49slopes} 
observed mass dependent inverse slopes ($T_{eff}$) of the $m_T$ spectra, and 
interpreted the trend as common production from an expanding hadron 
gas. The data were fitted with a collective transverse flow velocity 
and temperature of the gas at freeze-out, when the hadrons cease to 
undergo strong interactions. The hadron spectra are not, however, 
completely exponential in $m_T$ over a large range, and the deviation 
causes the extracted velocity and temperature parameters to depend 
upon the $m_T$ fit range.

We select published effective temperatures that were determined 
over a common fit range for all the data, $(m_T - m_0) < 1.2$ GeV, 
by an exponential fit (Equation ~\ref{eqn:mtexp}) to the 
measured $m_T$ distributions for each hadron spectrum produced 
at midrapidity.
\begin{equation}
\frac{dN}{m_T dm_T} = A e^{-m_T/T_{eff}}
\label{eqn:mtexp}
\end{equation}
The experiments include NA44 \cite{NA44flow,NA44kpi,NA44protons,NA44pPbPb} 
and WA97 \cite{WA97pbpb,WA97} at the SPS; and, at the ISR, Alper 
et al. \cite{alper} and Guettler et al. \cite{guettler}.  
For pions, the low-$p_T$ region of ($m_T-m_0) <0.3$ GeV, 
populated by decay of baryonic resonances, was 
systematically excluded.  The effective temperatures used 
are given in Table ~\ref{tab:slopes} with the references 
noted accordingly.
\begin{table*}
\caption{Inverse slope parameters (in MeV) of hadrons 
for p+p, p+nucleus, S+S, S+Pb, and Pb+Pb colliding systems at CERN
energies. The errors are statistical and systematic, respectively.}  
\begin{ruledtabular}
\begin{tabular}{c|ccccccc}  \hline
Hadron &Pb+Pb &S+Pb &S+S  &p+Pb  &p+S  &p+Be &p+p\\ \hline 
$\pi^+$ &156$\pm$6$\pm$23\footnotemark[1] 
&165$\pm$9$\pm$10\footnotemark[2]
&148$\pm$4$\pm$22\footnotemark[1] 
&145$\pm$3$\pm$10\footnotemark[2] 
&139$\pm$3$\pm$10\footnotemark[2]  
&148$\pm$3$\pm$10\footnotemark[2] 
&139$\pm$13$\pm$21\footnotemark[3]\\ 
$K^+$ 
&234$\pm$6$\pm$12\footnotemark[1]
&181$\pm$8$\pm$10\footnotemark[2]  
&180$\pm$8$\pm$9\footnotemark[1]
&172$\pm$9$\pm$10\footnotemark[2] 
&163$\pm$14$\pm$10\footnotemark[2] 
&154$\pm$8$\pm$10\footnotemark[2] 
&139$\pm$15$\pm$7\footnotemark[3] \\ 
p     
&289$\pm$7$\pm$14\footnotemark[4]
&256$\pm$4$\pm$10\footnotemark[5]  
&208$\pm$8$\pm$10\footnotemark[1]
&203$\pm$6$\pm$10\footnotemark[5] 
&175$\pm$30$\pm$10\footnotemark[5] 
&156$\pm$4$\pm$10\footnotemark[5] 
&148$\pm$20$\pm$7\footnotemark[3] \\ 
$\Lambda$ 
&289$\pm$8$\pm$29\footnotemark[6]
&---  &--- 
&203$\pm$9$\pm$20\footnotemark[7]
&---  &---  &--- \\ 
$\overline \Lambda$ 
&287$\pm$13$\pm$29\footnotemark[6]
&---   &---  
&180$\pm$15$\pm$18\footnotemark[7]
&--- &---  &---  \\ \hline
\end{tabular}
\label{tab:slopes}
\end{ruledtabular}
\footnotetext[1]{Reference \cite{NA44flow} (NA44 Collaboration).}
\footnotetext[2]{Reference \cite{NA44kpi} (NA44 Collaboration).}
\footnotetext[3]{Reference \cite{alper,guettler} (ISR).}
\footnotetext[4]{Reference \cite{NA44pPbPb} (NA44 Collaboration).}
\footnotetext[5]{Reference \cite{NA44protons} (NA44 Collaboration).}
\footnotetext[6]{Reference \cite{WA97pbpb} (WA97 Collaboration).}
\footnotetext[7]{Reference \cite{WA97} (WA97 Collaboration).}
\end{table*}

\begin{figure} [ht]
\resizebox{\columnwidth}{!}{\includegraphics{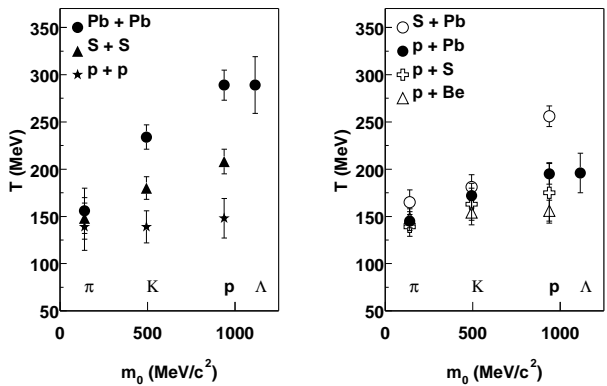}}
\caption[]{
The effective temperatures are plotted against particle mass for
positive particles produced in symmetric (A + A) collisions (left) and
asymmetric (p + A and A + B) collisions (right) at CERN energies.  Error
bars are statistical only.  In p+Pb (solid points), the effective 
temperature shows a slight dependence on particle mass.
}
\label{fig:fig1}
\end{figure}

For all but p+p and p+Be, the $\pi$, K, p and $\Lambda$ spectra 
were analyzed.  The mass dependence of $T_{eff}$ for positive 
particles is shown in Figure ~\ref{fig:fig1}; it is clear that a 
linear increase with particle mass exists for heavy colliding systems.

The dependence of $T_{eff}$ on both mass and the number of 
participating nucleons in the collision indicates radial 
expansion.  The effective temperature includes the local 
temperature of a small piece of matter and its 
collective velocity (for motion in two dimensions, the classical $T = 
m_{0}\langle v_{x} \rangle^{2}/2 + m_{0}\langle v_{y}\rangle^{2}/2 = m_{0}\langle v \rangle^{2}$, where $\langle v \rangle = 
\langle v_{x} \rangle = \langle v_{y} \rangle$ in circular coordinates 
in the plane transverse to the collision axis).  Simple exponential fits 
treat each particle spectrum as a thermal source, so the collective 
expansion velocity cannot be extracted reliably from a single particle 
spectrum; however, by using the information from all the particles, 
the expansion velocity can be inferred.  For example, NA44 
\cite{NA44flow} used
\begin{equation}
T_{eff} = T_{fo} + m_{0}\langle\beta_{t}\rangle^{2} \label{empirical}
\end{equation}
to separate the effective temperature into the two respective 
contributions (thermal and collective motion).  

We fit equation ~\ref{empirical} to the effective temperatures 
for particles produced in each collision system to determine 
the conditions at freeze-out.  Equation ~\ref{empirical} assumes 
that the flow velocity $\langle\beta_{t}\rangle$ is independent of 
particle mass $m_{0}$ (all particles collectively travel at the same 
radial velocity) \cite{csorgo}.  It also assumes that the pions, 
kaons, and protons have the same freeze-out temperature $T_{fo}$. 

Positive and negative hadrons yield equivalent freeze-out 
parameters. Omitting $\Lambda$ and $\overline\Lambda$ from 
the fits with Equation ~\ref{empirical} does not change the 
resulting parameters.  We use slopes from exponential fits to the 
measured $m_T$ spectra, as these describe the data well over the 
limited $m_T$ range 0.3 $ \le m_T-m_0 < $ 1.2 GeV. The simple relationship 
between $T_{eff}$, $T_{fo}$, and $\beta_t$ was shown to agree well with 
hydrodynamical fits to the same data \cite{NA44flow}, justifying the use 
of this simple form.

\begin{figure} [ht]
\resizebox{\columnwidth}{!}{\includegraphics{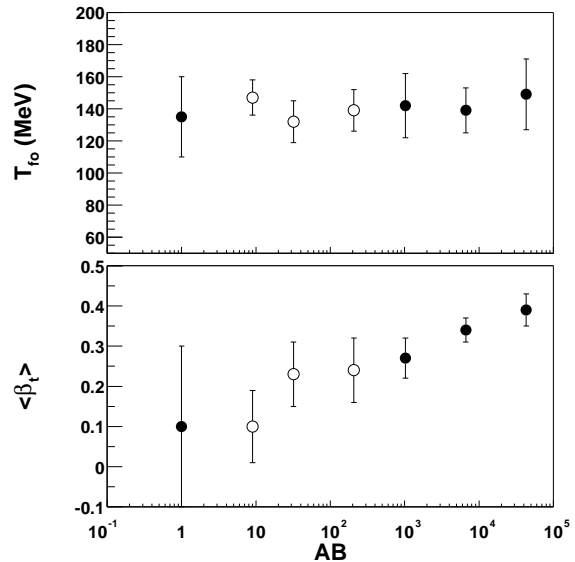}}
\caption[]{
The freeze-out temperature $T_{fo}$ (top) and mean radial expansion 
velocity $\langle\beta_t\rangle$ (bottom) plotted for each system 
in terms of system size AB at CERN energies.  The open symbols are 
p + A systems which follow the trend of increasing collective flow 
with system size.
}
\label{fig:fig2}
\end{figure}

Figure ~\ref{fig:fig2} shows the dependence of the parameters on the size 
of the colliding system.  The parameter $T_{fo}$ remains approximately 
140 MeV, regardless of system size; however, $\langle\beta_T\rangle$ 
increases, as would be expected from the larger pressure generated by 
copious particle production in the heavier colliding systems.  Our values 
of $T_{fo}$ and $\langle \beta_T \rangle$ are in reasonable agreement with 
other extractions of these parameters using two-particle correlations as 
well as single particle spectra.  It should be noted that 
Figure ~\ref{fig:fig2} includes p+A results as well as p+p and A+A 
collsions (symmetric and asymmetric).  The fact that p+Pb follows the 
trend of higher $\langle\beta_T\rangle$ for larger systems is remarkable, 
as significant collective flow in such collisions would not be 
expected. The effective temperatures most likely reflect the presence 
of initial state multiple parton scattering, often referred to as 
Cronin enhancement \cite{pa} already at relatively low hadron 
transverse momenta ($p_T \approx 1.3$ GeV/c for pions and $1.9$ GeV/c 
for protons).

In heavy ion collisions, initial state multiple parton scattering 
should also be present \cite{xnwcronin}.  At the relatively low 
$(m_T - m_0) <$ 1.2 GeV fitted here, we observe a much larger mass 
dependence of the slopes in A+A than in p+A collisions.  In this 
soft momentum range, one would expect the boost from collective 
expansion to dominate over multiple scattering at the parton level.  
At higher $m_T$, the relative contributions should be reversed.  
This can be checked by attributing all the velocity boost 
to collective flow, extrapolating the spectrum to higher momentum 
under this assumption, and comparing to data at higher $p_T$ or $m_T$.

In order to do this extrapolation, we assume that the soft 
physics processes in heavy ion collisions can be described by 
hydrodynamics.  The parameters in Figure 2 are used to extrapolate to 
higher transverse momentum the spectrum of pions from a hydrodynamically 
expanding gas.  We use a hydrodynamics parameterization from U. Heinz and 
coworkers, which assumes that all the particles decouple at a 
common freeze-out temperature $T_{fo}$, and that the source has 
an infinite length in the longitudinal direction 
(boost-invariance) \cite{schneder}. 

The transverse velocity profile of the source is parameterized 
as a function of the radius $\xi = \frac{r}{R}$, where R is 
the maximum radius of the expanding source at freeze-out, 
\begin{equation}
\beta_{t}(\xi) = \beta_{s}\xi^{n} . \label{prof}
\end{equation}
$\beta_{s}$ is the maximum surface velocity, and n ($= 1$ for linear) 
describes the profile.  A flat particle density distribution 
$f(\xi)$ $=$ const is also assumed.  The geometrical average 
of the transverse velocity is therefore given by
\begin{equation}
\langle\beta_{t}\rangle = \frac { \int{\beta_{s}\xi^{n}\xi d\xi}}{ \int{\xi d\xi} } =
\frac{2}{2+n} \beta_{s}. \label{geoave}
\end{equation}
Different velocity profiles, such as n $ = $ 1/2 or 2 result 
in lower or higher surface velocities, respectively.  If a 
different particle density distribution (for instance, a 
Gaussian) is used, then the average should be determined 
after weighting accordingly \cite{esumi}.  Using a linear 
velocity profile, the geometrical average flow 
velocity in the transverse plane is 
$\langle\beta_{t}\rangle$ $ = 2\beta_{t}/3$.

Each locally thermalized fluid element is given a transverse kick 
$\rho$ that depends on the radial position 
\begin{equation}
\rho = \tanh^{-1} \left( \beta_{t}\left( \xi \right) \right) . \label{rho}
\end{equation}
The transverse mass spectrum $\frac{dN}{m_T dm_T}$, of the particles is 
\begin{equation}
A \int_0^1{ m_{t} f(\xi) K_1 \left( \frac{ m_{t}\rm{cosh}(\rho) }{ T_{fo} } \right)
	I_0 \left( \frac{p_t\rm{sinh}(\rho)}{T_{fo}} \right) \xi d\xi} \label{hydro}
\end{equation}
which we integrate numerically.

\begin{figure*}
\includegraphics[scale=0.9,angle=0.0]{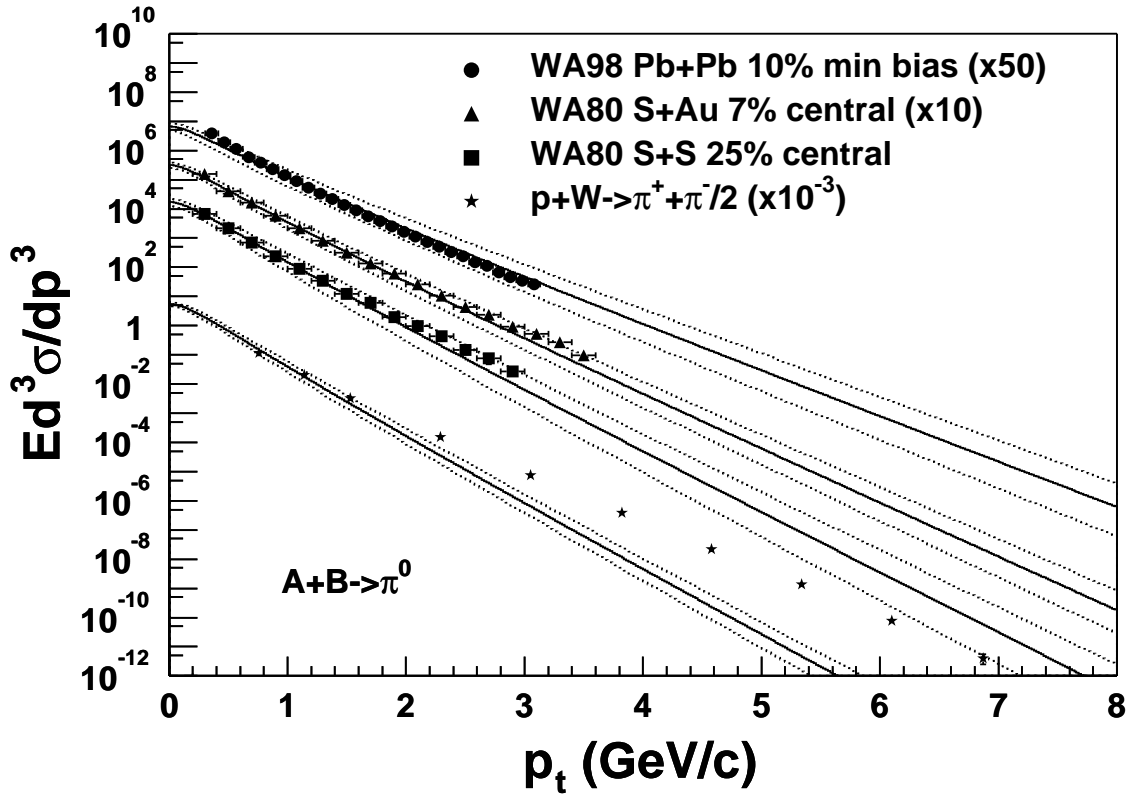}
\caption[]{
The hydrodynamically extrapolated pion spectrum compared to data 
to the highest available $p_T$ for multiple collision systems at CERN.
The dashed lines indicate error bands for each 
boosted spectrum.  The hydrodynamic distribution clearly deviates 
from the p+W data at $p_T >$ 2 GeV/c, while no such discrepancy 
is observed in the heavier colliding nuclei, illustrating the 
physics differences between p+A and A+A.
}
\label{fig:fig3}
\end{figure*}

We use Equation ~\ref{hydro} with the freeze-out temperature 
$T_{fo}$ and the maximum surface velocity $\beta_{t}$ from 
Figure ~\ref{fig:fig2} to extrapolate the particle spectra to high $p_T$.
  The normalization A is allowed to vary, as we compare only 
the shape of the extrapolated distribution to data.  
Figure ~\ref{fig:fig3} shows the resulting extrapolated spectra compared to 
the neutral pion spectra in Pb+Pb, S+Pb, 
S+S \cite{WA98spectra,WA80spectra} and charged pions in p+W collisions \cite{pa}.  (The equivalent p+Pb pion data in this momentum range was not 
available.)  The dashed lines are error bands, corresponding to uncertainties 
in $T_{fo}$ and $\beta_t$ from statistical and systematic errors 
in the effective temperatures for each collision system. 

For heavy ion collisions, the measured $\pi^0$ $p_T$ spectra are 
consistent with hydrodynamic distributions, though the highest 
$p_T$ point tends toward the edge of the upper error band.  
The moderate $p_T$ values measured to date in heavy ion collisions contain 
a mix of hard and soft processes. To cleanly observe hard processes 
in heavy ion collisions, measurements at $p_T \ge$ 4 GeV/c will 
be very valuable.

Pions produced in p+W collisions clearly deviate from the 
hydrodynamic parameterization at $p_T >$ 2 GeV/c, underscoring 
the differences in the physics between p+A and A+A collisions. The long tail 
in the p+W spectrum presumably reflects initial state parton 
multiple scattering and the onset of hard processes, which are 
masked by the contribution of boosted soft particles in heavy ion 
collisions.  By extrapolation, the p+A data may be used to 
estimate the contribution of multiple scattering of partons to 
the observed broadening of the hadron spectra in A+A collisions.

Stimulating discussions with Vladislav Pantuev, Xin-Nian Wang, 
Nu Xu, Axel Drees, R. Soltz, and S.C. Johnson are gratefully 
acknowledged.  This work was supported by the U.S. Department 
of Energy under grant No. DE-FG02-96ER40988.

\end{document}